\documentclass[12pt]{article}
\newcommand{\beq}{\begin{equation}}
\newcommand{\eeq}{\end{equation}}
\newcommand{\bea}{\begin{eqnarray}}
\newcommand{\eea}{\end{eqnarray}}

\newcommand{\gsim}{\lower.7ex\hbox{$
\;\stackrel{\textstyle>}{\sim}\;$}}
\newcommand{\lsim}{\lower.7ex\hbox{$
\;\stackrel{\textstyle<}{\sim}\;$}}

\newcommand{\eod}{\end{document}}

\begin{document}
\thispagestyle{empty}
\vspace*{-22mm}

\begin{flushright}
UND-HEP-12-BIG\hspace*{.08em}17

\end{flushright}
\vspace*{1.3mm}

\begin{center}
{\Large {\bf 
CP Violation in $\tau$ Decays at SuperB \& Super-Belle 
Experiments -- like Finding Signs of Dark Matter}}

\vspace*{10mm}

{ I.I.~Bigi$^a$}\\
\vspace{7mm}
$^a$  {\sl Department of Physics, University of Notre Dame du Lac, Notre Dame, IN 46556, USA}\\
{\sl email addresses: ibigi@nd.edu} \\

\vspace*{10mm}

{\bf Abstract}
\vspace*{-1.5mm}
\\
\end{center}
BaBar has found 
$A_{\rm CP}(\tau^- \to \nu K_S\pi^- [\geq \pi^0]) = (-0.36 \pm 0.23 \pm 0.11)\%$ with 2.8 $\sigma$ difference with SM prediction 
$A_{\rm CP}(\tau^- \to \nu K_S\pi^- ) = (0.36 \pm 0.01)\%$
based on $K^0 - \bar K^0$ oscillation. Four central points: 
(i) to establish both the `existence' of New Dynamics (ND) and its `features' in 
local CP asymmetries; 
(ii) to increase the number of final states to be probed;   
(iii) to emphasize correlations of CP violations between different final states; 
(iv) likewise for the correlations with $D^{\pm}$ final states.     
These measurements should be possible at SuperB \& Belle II experiments.

\vspace{3mm}

\hrule

\tableofcontents
\vspace{5mm}

\hrule

\section{Prologue: A New Era for `Our Universe'}
\label{ERA}

Since around the year 2000 we `know' that our Universe consists of $\sim$ 73 \% of 
`Dark Energy' (DE), 
$\sim$ 23 \% of `Dark Matter' (DM) and $\sim$ 4 \% of `Known Matter'. The existence of DE and 
DM is {\em hypothesized} to account for data about the patterns of our Universe. We have candidates for DM, but hardly any one for DE. 

The Standard Model (SM) has been very successful for describing known matter -- {\em except} for 
\begin{itemize}
\item 
neutrino oscillations  with 
$\theta_{12}, \theta_{23}, \theta_{13} > 0$ \cite{DAYA};  
\item 
huge asymmetry between matter vs. anti-matter -- i.e., our existence. 
\end{itemize} 
There is a good chance for leptonic dynamics producing `matter' vs. `anti-matter' asymmetry as a `shadow' effect of the `lepton' vs. `anti-lepton' asymmetry. Therefore we need New Dynamics (ND) in the interactions between known matter states. I am too old to think about DE, but  
make useful comments about the extistence, its features of ND and its correlations between flavour dynamics and DM. Before we said that DM is `somewhere overall' in the Universe. 
Now we have found one `road' of DM between two custers of known matter \cite{NATURE}.

After comments on E[W]DMs of electrons, $\mu$ and $\tau$ in Sect.\ref{EDM} 
I review CP asymmetries in $\tau$ decays in Sect.\ref{DCPV}, the landscapes 
\& theorical tools for finding ND in Sect.\ref{TOOL} and summarize searching for 
ND Sect.\ref{SUM1}. 

\section{Leptonic E[W]DMs}
\label{EDM}

Wonderful experimental research has lead to limits of $d_e < 1.1 \times 10^{-27}$ ecm from 
nuclear data. This is larger by several orders of ten than what SM can produce. 

Muon EDM has been probed leading to $d_{\mu} = (- 0.1 \pm 0.9   ) \times 10^{-19}$ ecm from 
PDG. Using a `simple scaling' from $d_e$ one gets $d_{\mu} < 2.5 \times 10^{-25}$ ecm; 
however suggested ND models make $d_{\mu} < 10^{-22}$ ecm. If future data give evidence of 
$d_{\mu} \sim 10^{-20} - 10^{-21}$ ecm, theorists will probably come up with ND models that can generate those values \cite{EDM}.  

PDG averaged data on 
$e^+e^- \to \tau^+\tau^-$ give Re[Im]$d_{\tau}^w < 0.50[1.1] \times 10^{-17}$ ecm. \\
My bet for finding ND in lepton forces underlaying matter-antimatter asymmetry: 
(i) Gold Medal: CP violation in neutrino oscillations; 
(ii) Silver Medal: EDMs; 
(iii) 
Bronze Medal: CP asymmetries in $\tau$ decays. 
Next week I might give the Gold Medal to EDMs and then change it again during next 5 - 10 years, before data can available. On the other hand for Austrian, German \& Swiss people getting a Bronze Medal is a great success.

\section{CP Asymmetries in $\tau$ Decays}
\label{DCPV}

Even if CP violation in charged leptons dynamics is not connected with matter-antimatter asymmetry, it is important to probe CP symmetry in $\tau$ decays with high accuracy: 
at the level of ${\cal O}(0.1 \%)$ in $\tau \to \nu [K\pi/K2\pi]$ might have 
roughly the same sensitivity of ND in the amplitude as searching for BR$(\tau \to \mu \gamma)$ at the level of $10^{-8}$ \cite{CPBOOK}. For CP odd observables in a SM allowed decay are {\em linear} in a ND amplitude, while in SM forbidden ones the rates are {\em quadratic} in ND amplitudes: 
CP odd $\propto T^*_{\rm SM} T_{\rm ND}$ vs. LFV $\propto |T_{\rm ND} |^2$.

BaBar Coll. have produced data with finding CP violation in $\tau$ decays is  
a `hope' \cite{BABARTAU}: 
\beq 
A_{\rm CP}(\tau^- \to \nu K_S\pi^- [\geq \pi^0] ) = (-0.36 \pm 0.23 \pm 0.11)\%  \; . 
\eeq 
CP violation established in $K^0 - \bar K^0$ oscillations gives as predicted \cite{BSTAU,NIRTAU}: 
\beq 
A_{\rm CP}(\tau^- \to \nu K_S\pi^- )= 2 {\rm Re}\, \epsilon_K = (0.36 \pm 0.01)\% \; ;   
\eeq 
i.e., there is a difference of 2.8 sigma between these two values. Thus there is experimental sign of global CP violation in $\tau$ decays. Furthermore SM gives 
\beq 
A_{\rm CP}(\tau^- \to \nu K_S[\pi^- \& K^-] [\geq \pi^0] )_{\rm SM}= 2 {\rm Re}\, \epsilon_K 
\; .  
\eeq

\subsection{General Comments}

{\em Global} asymmetries are often much more suppressed than local ones. 
One needs to probe different final states -- include 
three- and four-body ones to established its   
{\em existence} of ND. Furthermore it is crucial to determine its {\em features}. 
We should  
focus on transitions that are CKM suppressed in SM -- like $\tau \to \nu [K\pi's] $ 
(or even better $\tau \to \nu [K\eta \pi's]$), when one has enough data in the future -- where one has a good 
chance to identify both the impact and features of ND with less `background' from SM amplitudes. 

One needs conceptual lessons for probing CP violations in 
$\tau^- \to \nu [Kh_i ]^- / \nu [Kh_ih_j]^- (h = \pi, \eta /\nu [3K]^- /$ $\nu [K3\pi]^- $ 
{\em separately} to understand the underlaying dynamics. 
First one compares  
the $\tau^-$ and $\tau^+$ widths of these final states (FS); however one should measure 
`local' asymmetries as defined later.  
 
Ignoring {\em weak} final state interactions (FSI) CPT invariance predicts 
\bea
\Gamma (\tau ^- \to \nu X_{S=-1}) &=& \Gamma (\tau ^+ \to \bar \nu X_{S= 1}) \\
\Gamma (\tau ^- \to \nu X_{S=0}) &=& \Gamma (\tau ^+ \to \bar \nu \bar X_{S=0})
\eea
with $X_{S=-1} = \bar K^0 \pi^-/K^- \pi^0[\eta]/\bar K^0 \pi^-\pi^0[\eta]/K^-\pi^+\pi^-/
K^-\pi^0\pi^0/K^-K^+K^-/K^-\bar K^0 K^0/ 
\\ \bar K^0(3\pi)^-/K^-(3\pi)^0$ and 
$X_{S=0} = \pi^-\pi^0/\pi^-\eta/4\pi/3\pi/K\bar K/\pi K \bar K/2\pi K\bar K$ etc.  

Other symmetries and their violations connect same FS on different scales. 
In particular $\tau^- \to \nu \pi^-\pi^0[\eta]$ can combine with 
$\tau^- \to \nu 4\pi/\nu 2\pi K \bar K$ to get close to CPT symmetry and for 
$\tau^- \to \nu 3\pi$ with $\tau^- \to \nu \pi K \bar K$. 
\\
Three items have to be dealt with:
\\ 
(1) 
One measures FS with $K_S$, $K_L$ and their interferences. 
$K^0 - \bar K^0$ 
oscillation impacts CP asymmetries as expressed by 2Re$\, {\epsilon_K }$ in a {\em global} way for channels. 
\\
(2) 
Mixing between $\bar K^0 \pi^- \Leftrightarrow K^- \pi^0$, 
$\bar K^0 \pi^0 \Leftrightarrow K^- \pi^+$ and  
$K^- K^+ \Leftrightarrow K^0 \bar K^0$ happen by FSI -- like it does  
for $K\pi \leftrightarrow K\eta$, but on reduced level. 
\\
(3) 
Theorerical tools exist for $\pi\pi$, $\pi K$, $K\bar K$ 
non-perturbative interactions based on dispersion relations and others that use data 
in different ways. 

One can measure rates and CP asymmetries in $\tau ^- \to \nu [K\pi's]^- $ vs. 
$\tau ^+ \to \bar \nu [K \pi's]^+$  
and to calibrate ratios of  $\tau ^- \to \nu [\pi's]^- $ vs. $\tau ^+ \to \bar \nu [\pi's]^+$, where one expects that even 
ND can hardly produce measurable asymmetries.

For $\tau \to \nu [K2\pi]/ \nu [3K] /\nu [K3\pi] $ one has more CP odd observables through {\em moments} and their {\em distributions} to check the impact of ND. Those are described by 
total four- \& five-body  FS -- and therefore {\em hadronic} three- \& four-body FS  with {\em distributions} of hadrons. There are several theoretical technologies \cite{CPBOOK,MIRANDA,MIKE}.

Unless one has 
{\em longitudinally polarized} $\tau$, one needs differences in both the weak and strong phases to 
generate CP asymmetries in $\tau \to \nu [K\pi]$. Non-zero T odd observables can be produced by 
FSI without CP violation. On the other hand true CP asymmetries can be probed for $\tau^-$ vs. $\tau ^+$ decays.

Finding CP asymmetries in $\tau$ decays (beyond CP violation in $K^0 - \bar K^0$ oscillatings)  is a clear evidence for impact of ND. However one has to be prepared for very small effects and to depend on correlations with different FS.

There are important three points:  
(i) For expected data from SuperB and Belle II (and even for existing 
archives) one has to proceed step-by-step in experimental and theoretical work.  
(ii) One has to probe FS with $\eta$.  
(iii) One has to measure CP asymmetries 
in the pair of $\tau^-\tau^+$ to `tag' them by $[e^+\nu \bar \nu]f_{\tau^-}$ and 
$[\mu^+\nu \bar \nu]f_{\tau^-}$.

SuperB experiment could produce a 
pair of longitudinally polarized $\tau$ and therefore probe T {\em odd} moments and their distributions in $\tau \to \nu h_1h_2/\nu h_1h_2h_3/\nu h_1h_2h_3h_4$ decays.

\subsection{Landscape for $\tau$ Decays}
\label{LAND}
The complex FS can be probed with classes of CP {\em odd} observables.  
\\ 
(i) 
{\em 0-dimensional} observables: $\Gamma (\tau ^- \to \nu [\bar K h_i]^-)$, 
$\Gamma (\tau ^- \to \nu [\bar Kh_ih_j]^-)$ etc. or the 
{\em averaged} value of angles between planes of $\nu-K$ and $h_i-h_j$ etc. with $h_i = \pi , \eta$.  \\
(ii)
{\em 1-dimensional} ones: {\em lines} in $\tau$ rest frame like 
$\frac{d}{dE_{\nu}}\Gamma (\tau ^- \to \nu [Kh_i]^-)$, 
$\frac{d}{dE_{\nu}}\Gamma (\tau ^- \to \nu [K2h_ih_j]^-)$ etc. or {\em angles} between planes of $\nu-K$ and $h_i-h_j$ in $\tau^- \to \nu [Kh_ih_j]^-$ etc.\\
(iii) 
{\em 2-dimensional} ones: {\em patterns} in `averaged Dalitz plots' in  
$\tau ^- \to \nu X^-_{s\bar u}$. \\
(iv)  
{\em 3-dimensional} ones etc. 
 \\
Three-, four- and five-body FS have two, three and four hadrons $\pi$, $K$ and/or $\eta$. One can probe the complex FS with independent of $\tau$ 
production asymmetry \cite{PICH,DEL,DATTA,CLEO}.  

Belle/ BaBar and in the future SuperB/Belle II give the best landscape to 
probe $\tau^-\tau^+$ pair and thus to measure correlations. In  
Ref.\cite{CPBOOK} it was mentioned that 
\beq
e^+e^- \to \tau ^+ \tau^- \to [l^+\nu \bar \nu]_{\tau^+} \nu f_{\tau^-} \; \; {\rm vs.} \; \; 
[l^-\nu \bar \nu]_{\tau^-} \bar \nu \bar f_{\tau^+} \; , \; l=e,\mu 
\eeq 
can be probed T {\em odd} observables $\vec p_{l^+}\cdot (\vec p_{h_1}\times \vec p_{h_2})$
vs. $\vec p_{l^-}\cdot (\vec p_{\bar h_1}\times \vec p_{\bar h_2})$ and their moments.

Transfer of longitudinally polarized $e^+e^-$ to longitudinally ones $\tau$ will 
produce more observables in CP asymmetries depending on the angle between $e^+e^-$ 
and $\tau^+\tau^-$ pairs.

\section{Models for ND in $\tau$ Decays and Tools}
\label{TOOL}

There are several classes of ND models: 
\begin{itemize}
\item 
The `natural' ones are based on charged Higgs exchanges, 
in particular for $X_{S\neq 0}$ FS, since SM amplitudes are relatively suppressed 
compared to $S = 0$ ones.
\item 
$W_L - W_R$ mixing affects 
$\tau ^- \to \nu X^-_{S=-1}$ vs. $\tau ^+ \to \bar \nu X^+_{S=1}$ decays. 
Limits one gets  
from $B$ and $K$ transitions depend on sizable inputs from theoretical uncertainties. 
CP asymmetries are produced by the phase between SM $W_L$ and $W_L-W_R$ mixing amplitudes; 
therefore probing them offer higher sensitivity for ND. 
\item 
`New' $W_L$  bosons can couple with quark 
and leptons differently than SM $W_L$. One can compare the ratios of 
$\tau^- \to \nu X_h^-$/ $\tau^- \to l^-\bar \nu_l \nu_{\tau }$  vs.  
$\tau^+ \to \bar \nu \bar X^+_h$/ $\tau^+ \to l^-\bar \nu_l \nu_{\tau }$.
\item 
An exotic option is the class of leptoquark exchanges  
for $\tau^- \to [\bar u s] \nu$. 
\end{itemize}

\subsection{Correlations of $\tau ^{\mp}$ with $D^{\mp}$ Decays}

One should not stop after probing hadronic two-body FS. One needs to find the existence 
and the features of ND by measuring {\em many}-body FS with accuracy -- namely for 
{\em hadronic three- and four-body} FS: 
\\
(i) $D^- \to K_S\pi^-/K^- \pi^0/K^- \eta$ vs.    
$\tau ^- \to \nu K_S\pi^-/\nu K^- \pi^0/ \nu K^- \eta$; 
\\
(ii)   
$D^- \to K_S\pi^-\pi^0/ K_S\pi^-\pi^0 /K^- \pi^+\pi^-/K^- \pi^0 \pi^0$  vs.      
$\tau ^- \to \nu K_S\pi^-\pi^0/\nu K_S\pi^-\eta /\nu K^- \pi^+ \pi^-/$ $\nu K^- \pi^0\pi^0$; 
\\
(iii)  
$D^- \to K^-\pi^+\pi^-\pi^0/K_S\pi^-\pi^-\pi^+$ vs. 
$\tau^- \to \nu K^- \pi^+\pi^-\pi^0/\nu K_S\pi^-\pi^-\pi^+$;
\\
(iv) $D^- \to K_S K^-$ vs.    
$\tau ^- \to \nu K_S K^-$;  
\\
(v)  
 $D^- \to K_S K^-\pi^0/K_SK_S\pi^-$   vs.    
$\tau ^- \to \nu K^- K_S\pi^0/  \nu K_SK_S\pi^-$;  
\\
(vi)  
$D^- \to K^-K_S\pi^+\pi^-/K^+K_S\pi^-\pi^-$ vs. 
$\tau^- \to \nu K^-K_S\pi^+\pi^-/ \nu K^+K_S\pi^-\pi^-$ etc. 

The landscapes of multi-body FS for $D^{\pm}$ and 
$\tau ^{\pm}$ decays are both different and similar:  
\begin{itemize}
\item
One has Cabibb favoured 
$\tau^- \to \nu d\bar u + q \bar q$ and singly suppressed  
$\tau^- \to \nu s\bar u + q \bar q$. The SM produces no CP asymmetries 
in $\tau ^- \to \nu K^- [\pi 's]^0/\nu K^-K_SK_S/K^-K^+K^-/ \\ \nu K_SK_S[\pi 's]^-$ and   
in $\tau ^- \to \nu K_S [\pi 's]^-/\nu K^-K_S[\pi 's]^0$ global one due to $K^0 - \bar K^0$ oscillations by 2Re$\, \epsilon_K$.  

\item
The landscape for $D^{\mp}$ decays is complex for several reasons.  
\begin{itemize}
\item 
DCS $D^- \to K^- X_{S=0}^0$ give no CP asymmetries in SM. 

\item 
Again SCS $D^- \to K^- K_S X_{S=0}^0$ have input on global CP asymmetry from 
$K^0 - \bar K^0$ oscillation by 2Re$\, \epsilon_K$. 

\item
Furthermore SM produces direct CP violation in $D^- \to K^-K_S\pi^0/K^+K^-\pi^-/$
$\pi^+\pi^-\pi^-/\pi^-\pi^0\pi^0/K^-K_S\pi^+\pi^-/K^+K^-\pi^-\pi^0/
\pi^-\pi^-\pi^+\pi^0$ (ignoring FS with $\eta$). 

\item 
Both SM and ND affect the {\em topologies} of three- and four-body FS. 
\item  
FS with $K_S$ are affected by DCS and CF amplitudes. However the 
interferences in SM give a small corrections to $K^0 - \bar K^0$ oscillations 
\cite{CPBOOK}. 
\end{itemize}

\end{itemize}
There is little reason why ND should affect $D^-$ and $\tau^-$ decays in the 
same way.

\subsection{Tools and Technologies}

Heavy flavour states -- namely $H_b$, $H_c$ and $\tau$ -- have many multi-body FS, 
and they offer more observables. 
One will not have infinite 
data; therefore we have to be practical and go step-by-step to probe CP asymmetries. 

Usually one focuses on hadronic (quasi-)two-body FS in $H_b$ and $H_c$ transitions for 
experimental and theoretical reasons. For beauty transitions we know that the SM generates 
at least the leading source of CP violation; therefore we have to search for input from 
ND for nonleading source(s).  
For charm transitions SM might produce nonleading CP asymmeries; however ND should 
gives us leading ones, but still only small ones. In both cases I want to know not only the 
`existence', but also its `features'. Therefore one has to probe CP asymmetries with high 
accuracy. Three-(\& four-)body FS needs more data and `working' -- but also gives us more 
lessons about the underlaying dynamics. It makes also 
reason to use different `theoretical technologies' for probing CP asymmetries 
`locally' in the FS:  
\\
(a) customary {\em fractional} asymmetry 
$\Delta (i) \equiv \frac{N(i) - \bar N(i)}{N(i) + \bar N(i)}$;
\\
(b) 
`Miranda Procedure' \cite{MIRANDA} based on analyzing the {\em significance}  
$\Sigma (i) \equiv \frac{N(i) - \bar N(i)}{\sqrt{N(i) + \bar N(i)}}$,  
which are powerful for finding CP asymmetries and to localize them;  
\\ 
(c) another one has been suggested in Ref.\cite{MIKE} based on unbinned multivariante results. More will probably come encouraged by future LHCb data and their interpretations.

\section{Summary}
\label{SUM1}

SM cannot generate measurable CP asymmetries in $\tau ^- \to \nu [K^- \pi's]$ and  
a value of $(0.36 \pm 0.01) \%$ in widths for $\tau ^- \to \nu [K_S \pi's]$. ND (like 
with charged Higgs exchanges) can affect 
these decays with hadronic two-, three- and four-body final states significantly with probing 
regions of interference between different resonances. To be more precise: 
\begin{itemize}
\item 
One has to measure $A_{\rm CP} (\tau^- \to \nu [K\pi ]^-)$, 
$A_{\rm CP} (\tau^- \to \nu [K 2\pi ]^-)$, 
$A_{\rm CP} (\tau^- \to \nu [3K ]^-)$ and 
$A_{\rm CP} (\tau^- \to \nu [K3\pi ]^-)$. 

\item 
As emphasized before about $B$ and $D$ decays with three- and four-body final states, one 
gets contributions from resonances and their interferences for CP asymmetries. 
However `global' asymmetries averaged over the total widths are significantly smaller than 
individual contributions.  
\item 
Therefore it is very important to probe the `topologies' in the Dalitz plots. 

\item 
For $\tau^- \to \nu [K\pi ]^-$ one can probe interference between vector and scalar states, 
which are somewhat suppressed. 
For $\tau^- \to \nu [K2\pi ]^-/[3K]^-$ one can probe T odd moments due to vector and axial vectors 
exchanges and even more for $\tau^- \to \nu [K3\pi ]^-$, which should not be suppressed in general. 

\item 
On the first step to probe the final states as discussed above one can look for {\em local} asymmetries in 
$\tau ^- \to \nu [3K + K2\pi ]^-$ vs. $\tau ^+ \to \bar \nu [3K + K2\pi ]^+$.

\end{itemize}
SuperB and Belle II experiments should be able to probe the {\em whole} area of 
$\tau \to \nu [K\pi/K2\pi/ \\ 
3K/K3\pi]$ transitions with {\em neutral} pions in the final states. 

One more comment about CP asymmetries in $\tau$ decays: These comments about the  
impact of ND is focused on 
semi-hadronic $\tau$ transitions. It is most likely to affect also $B$ and $D$ decays, but it 
could `hide' more easily there due to larger effects (in particular for $B$ transitions) and less control over non-perturbative QCD effects.  

One has to probe the distributions of FS in $\tau$ (and $B/D$) decays to 
find the impacts and the features of ND(s) based on `binned' \cite{MIRANDA} and 
`unbinned multivariate' \cite{MIKE} results.  

Final recent lesson from molecular biology about information in DNA : It was thought 
most DNA have `junk’ informations, no reason to probe it. 
Now experts say that at least 80 \% of DNA are active and needed to understand informations.  
\\
Analogy: {\em Most features of ND are probed in multi-body FS in $\tau$, $H_b$ and 
$H_c$ decays}.

\vspace{0.5cm}

{\bf Acknowledgments:} I are very thankful for the organizers for this wonderful workshop 
at Nagoya. This work was supported by the NSF under the grants numbers PHY-0807959 and PHY-1215979. 

\vspace{4mm}


\end{document}